\documentclass[10pt,conference]{IEEEtran}
\usepackage{epsfig}
\usepackage{amssymb,amsmath}
\usepackage{graphicx}
\usepackage{epic}
\usepackage{color}

\newcommand{\openone}{\leavevmode\hbox{\small1\normalsize\kern-.33em1}}

\newcommand{\dl}{{\mathtt l}}
\newcommand{\dr}{{\mathtt r}}

\newcommand {\beq} {\begin{equation}}
\newcommand {\eeq} {\end{equation}}
\DeclareMathOperator{\expectation}{\ensuremath{\mathbb{E}}} 
\DeclareMathOperator{\encoder}{\ensuremath{f}} 
\DeclareMathOperator{\distortion}{\ensuremath{d}} 
\DeclareMathOperator{\decoder}{\ensuremath{g}} 

\newtheorem{theorem}{Theorem}{}
{}
{}
{}
{}
\newtheorem{example}{Example}{}

\DeclareMathOperator{\naturals}{\ensuremath{\mathbb{N}}}
\newcommand {\GF}{\mathbb{F}_2}

\newcommand {\code} {\hat{\mathcal S}}

\newcommand {\prob} {{\mathbb P}}

\begin{document}

\title{Lower Bounds on the Rate-Distortion Function  of Individual LDGM Codes}

\author{
\authorblockN{Shrinivas Kudekar and R{\"u}diger Urbanke}
\authorblockA{EPFL, School of Computer and Communication Sciences, Lausanne 1015, Switzerland}
}
%

\maketitle

\begin{abstract} 
We consider lossy compression of a binary symmetric source by means of
a low-density generator-matrix code.  We derive two lower bounds on the rate
distortion function which are valid for any low-density generator-matrix
code with a given node degree distribution $L(x)$ on the set of generators
and for any encoding algorithm.  These bounds show that, due to the
sparseness of the code, the performance is strictly bounded away
from the Shannon rate-distortion function. In this sense, our bounds
represent a natural generalization of Gallager's bound on the maximum
rate at which low-density parity-check codes can be used for reliable
transmission.  Our bounds are similar in spirit to the technique recently
developed by Dimakis, Wainwright, and Ramchandran, but they apply to {\em
individual} codes.
\end{abstract}

\section{Introduction} 
\begin{figure}[htp]
\begin{center}
\input{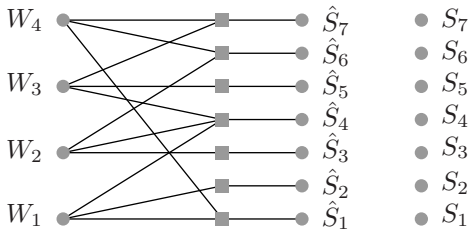}
\end{center}
\caption{\label{fig:ldgmtanner}
The Tanner graph corresponding to a simple LDGM code used for
lossy compression of a BSS. We have $m=7$, $R=\frac47$, 
and $L(x)=x^3$.
}
\end{figure}

We consider lossy compression of a binary symmetric source
(BSS) using a low-density generator-matrix (LDGM) code as shown in
Figure~\ref{fig:ldgmtanner}. More precisely, let $S \in \GF^m$ represent
the binary source of length $m$.  We have $S=\{S_1,S_2,\dots,S_m\}$,
where the $\{S_i\}_{i=1}^{m}$ are iid random variables with
$\prob\{S_i=1\}=\frac12$, $i\in [m]$.  Let $\mathcal{S}$ denote the set
of all source words.

Given a source word $s \in {\mathcal S}$, we compress it by mapping it to
one of the $2^{m R}$ index words $w \in {\mathcal W} = \GF^{m R}$, where
$R$ is the {\em rate}, $R \in [0, 1]$. 
We denote this encoding map by
$\encoder: s \mapsto W$ (the map can be random).  The reconstruction is
done via an LDGM code determined by a sparse binary $m R \times m$ generator
matrix $G$.  Let $\hat{s}$ denote the reconstructed word associated to
$w$.  We have $\hat{s} = w G$. We denote this decoding map by $\decoder:
w \mapsto \hat{s}$.  Let $\code$ denote the code, $\code=\{\hat{s}^{(1)}, \dots,
\hat{s}^{(2^{m R})}\}$, $\hat{s}^{(i)} \in \GF^m$. The codewords are not necessarily distinct.

We call the components of the index word $w=\{w_1, \dots, w_{m R}\}$
the {\em generators} and the associated nodes in the factor graph
representing the LDGM code the {\em generator nodes}. 
We assume that these generators nodes have a normalized degree distribution 
$L(x)=\sum_{i} L_i x^i$. This means that $L_i$ represents the fraction 
(out of $m R$) of generator nodes of degree $i$.

We are interested in the trade-off between rate and distortion which is achievable 
in this setting. Let $\distortion(\cdot, \cdot)$ denote the Hamming distortion function,
$\distortion: \GF^m \times \GF^m \rightarrow \naturals$.
The average distortion is then given by
\begin{align*}
\frac1{m} \expectation[d(S, \decoder(\encoder(S))].
\end{align*}
We are interested in the minimum of this average distortion, where the minimum
is taken over all LDGM codes of a given rate, generator degree distribution $L(x)$, and length,
as well as over all encoding functions.

\section{Review}
Given the success of sparse graph codes applied to the channel coding problem, it
is not surprising that there is also interest in the use
of sparse graph codes for the source coding problem.
Martinian and Yedidia \cite{MaYe03} were probably the first to 
work on lossy compression using sparse graph codes. 
They considered a memoryless ternary source with erasures and demonstrated a duality result between
compression of this source and the transmission problem over
a binary erasure channel (both using iterative encoding/decoding).
Mezard, Zecchina, and Ciliberti  \cite{CiMe05} considered the lossy compression
of the BSS using LDGM codes with a Poisson distribution on the generators.
They derived the one-step replica symmetry-breaking (1RSB) solution 
and the average
rate-distortion function. According to this analysis, this ensemble approaches
the Shannon rate-distortion curve exponentially fast in the average degree.
They observed that the iterative interpretation associated to the 1RSB analysis
gives rise to an algorithm, which they called {\em survey propagation}. In
\cite{CiMeZe05} the same authors implement an encoder that utilizes a Tanner graph with random non-linear
functions at the check nodes and a {\em survey propagation} based
decimation algorithm for data compression of the BSS.  
In \cite{WaM05}, Wainwright and Maneva also considered the lossy compression of a
BSS using an LDGM code with a given degree distribution. They showed how survey
propagation can be interpreted as belief propagation algorithm 
(as did Braunstein and Zecchina \cite{BZ03})
on an enlarged set of assignments and demonstrated
that the survey propagation algorithm  is a practical and efficient encoding
scheme.
Recently, Filler and Friedrich \cite{FiFr07} demonstrated experimentally that
even standard belief propagation based decimation algorithms using optimized
degree distributions for LDGM codes and a proper initialization of the messages
can achieve a rate-distortion trade-off very close to the Shannon bound. 
Martinian and Wainwright \cite{MaWa06,MaW06a,MaW06b} constructed {\em compound LDPC
and LDGM code ensembles} and gave rigorous {\em upper bounds} on their distortion
performance. A standard LDGM code ensemble is a special case of their
construction, hence they also provide {\em upper bounds} on the rate-distortion
function of LDGM ensembles. By using the first and second moment method they proved
that a code chosen randomly from the {\em compound ensemble} under optimal encoding and decoding achieves the Shannon
rate-distortion curve with high probability. Finally, they pointed out that such constructions are
useful also in a more general context (e.g., the Wyner-Ziv or the Gelfand-Pinsker
problem).
Dimakis et al \cite{DiWaRa07} were the first authors to provide
rigorous {\em lower bounds} on the rate-distortion function of LDGM code
ensembles. 
\begin{theorem}[Dimakis, Wainwright, Ramchandran \cite{DiWaRa07}]
\label{the:dwrbound} Let $\code$ be a binary code of blocklength $m$ and rate
$R$ chosen uniformly at random from an ensemble of left Poisson LDGM Codes with check-node degree
$\dr$. Suppose that we perform MAP decoding. With high probability the 
rate-distortion pair ($R,D$) achieved by $\code$ fulfills
\begin{align*} 
R & \geq \frac{1-h(D)}{1-e^{-\frac{(1-D)\dr}{R}}} > 1-h(D).
\end{align*}
\end{theorem}

\subsection{Outline}
In the spirit of Gallager's information theoretic bound for LDPC codes,
we are interested in deriving lower bounds on the rate-distortion function 
which are valid for {\em any} LDGM code with a given generator node degree distribution $L(x)$.
Our approach is very simple.
Pick a parameter $D$, $D \in [0, \frac12]$ (think of this parameter as
the distortion).  Consider the set of ``covered'' sequences \begin{align}
\label{equ:cofd} {\mathcal C}(D) & = \bigcup_{\hat{s} \in \code} {\mathcal
B}(\hat{s}, D m), \end{align} where ${\mathcal B}(x, i)$, $x \in \GF^m$,
$i \in [m]$, is the Hamming ball of radius $i$ centered at $x$. In words,
${\mathcal C}(D)$ represents the set of all those source sequences that
are within Hamming distance at most $D m$ from at least one code word.

Recall that for any $s \in {\cal S}$, ${\encoder}(s) \in {\mathcal W}$
represents the index word and that ${\decoder}({\encoder}(s))$ denotes the
reconstructed word. We have
\begin{align*}
\distortion(s, {\decoder}({\encoder}(s))) & \geq 
\begin{cases}
0, & s \in {\mathcal C}(D), \\
Dm, & s \in \GF^m \setminus {\mathcal C}(D).
\end{cases}
\end{align*}
Therefore,
\begin{align} 
&\frac1{m} \expectation[\distortion(S, {\decoder}({\encoder}(S)))]  \nonumber \\
& = \frac1{m}\sum_{s \in \GF^m}  2^{-m} \distortion(s, {\decoder}({\encoder}(s)))
 \geq \frac{2^{-m}}{m} \sum_{s \in \GF^m \setminus {\mathcal C}(D)} \distortion(s, {\decoder}({\encoder}(s))) \nonumber \\
& \geq 2^{-m} D |\GF^m \setminus {\mathcal C}(D)| \geq D \bigl(1-2^{-m} |{\mathcal C}(D)| \bigr).
\label{equ:averagedistortion}
\end{align}
If the codewords are well spread out then we know from Shannon's random coding
argument that for a choice $D=h^{-1}(1-R)$, $|{\mathcal C}(D)| \approx 2^m$, \cite{CoT91}.
But the codewords of an LDGM code are clustered since changing a
single generator symbol only changes a constant number of symbols in
the codeword.  There is therefore substantial overlap of the balls.
We will show that there
exists a $D$ which is strictly larger than the distortion corresponding
to Shannon's rate-distortion bound so that $|{\mathcal C}(D)|$ is exponentially
small compared to $2^m$ regardless of the specific code. From
(\ref{equ:averagedistortion}) this implies that the distortion is at
least $D$.

To derive the required upper bound on $|{\mathcal C}(D)|$ we use two
different techniques.  In Section~\ref{sec:boundviacounting} we use a
simple combinatorial argument.  In Section~\ref{sec:boundviatestchannel},
on the other hand, we employ a probabilistic argument based on the
``test channel'' which is typically used to show the achievability of
the Shannon rate-distortion function.

Although both bounds prove that the rate-distortion function is
strictly bounded away from the Shannon rate-distortion function for
the whole range of rates and any LDGM code, we conjecture that a
stronger bound is valid. We pose our conjecture as an open problem in
Section~\ref{sec:openquestions}.

\section{Bound Via Counting}\label{sec:boundviacounting}
\begin{figure}[htp]
\begin{center}
\input{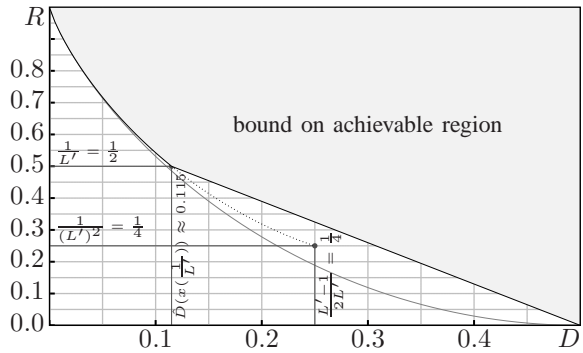}
\end{center}
\caption{\label{fig:rdconstruction}
Construction of the bound for codes with $L(x)=x^2$ so that $L'=2$ (all generator
nodes have degree $2$). The solid gray curve corresponds to the Shannon rate-distortion curve.
The black curve just above, which is partially solid and partially dotted, corresponds
to the curve $(\hat{D}(x), \hat{R}(x))$ for $x \in [0, 1]$. It starts at the point $(0, 1)$
(which corresponds to $x=0$) and ends at $(\frac{L'-1}{2 L'}=\frac14, \frac{1}{(L')^2}=\frac14)$ 
which corresponds to $x=1$. The straight line goes from the point $(\hat{D}(x(\frac{1}{L'})), \frac{1}{L'})$ to the point $(\frac12, 0)$. Any achievable $(R, D)$ pair must lie in the lightly shaded region.
This region is strictly bounded away from the Shannon rate-distortion function over the whole range.
}
\end{figure}

\begin{theorem}[Bound Via Counting]\label{the:boundviacounting}
Let $\code$ be an LDGM code with blocklength $m$ and with generator
node degree distribution $L(x)$ and define $L'=L'(1)$.  Let
\begin{align*}
f(x)  = \prod_{i=0}^{d} (1+x^i)^{L_i}, \;\;
a(x)  = \prod_{i=0}^{d} i L_i \frac{x^i}{1+x^i}, \\
\hat{R}(x)  = \frac{1-h(\frac{x}{1+x})}{1-\log \frac{f(x)}{x^{a(x)}}}, \;\;
\hat{D}(x)  = \frac{x}{1+x} - a(x) \hat{R}(x).
\end{align*}
For $R \in [\frac{1}{L'}, 1]$ let $x(R)$ be the unique positive solution of $\hat{R}(x)=R$.
Define the curve $D(R)$ as
\begin{align*}
& \begin{cases}
\frac12 \Bigl(1-R L' \bigl(1-2\bigl( \frac{x(\frac{1}{L'})}{1+x(\frac{1}{L'})} -
\frac{a(x(\frac{1}{L'}))}{\dl}\bigr)\bigr)\Bigr), &
 R \in [0, \frac{1}{L'}], \\
\hat{D}(x(R)), R \in [\frac{1}{L'}, 1].
\end{cases}
\end{align*}
Then, for any blocklength $m$, the achievable distortion of an LDGM code of rate $R$ and generator degree distribution $L(x)$
is lower bounded by $D(R)$.
\end{theorem}
Discussion: 
(i) As stated above, if we are considering a single code of rate $R$ then
the lower bound on the distortion is $D(R)$. If, on the other hand we are considering
a family of codes, all with the same generator degree distribution $L(x)$ but with
different rates $R$, then it is more convenient to plot the lower bound in a parametric
form. First plot the curve $(\hat{D}(x), \hat{R}(x))$ for $x \in [0, 1]$. Then connect
the point $(D=\frac12, R=0)$ to the point on the $(\hat{D}(x), \hat{R}(x))$ curve
with $\hat{R}(x)=\frac{1}{L'}$ by a straight line. The resulting upper envelope
gives the stated lower bound for the whole range. This construction is shown in 
Figure~\ref{fig:rdconstruction}.
(ii) 
Although this is difficult to glance from the expressions, we will
see in the proof that for any bounded generator degree distribution
$L(x)$ the performance is strictly bounded away from the Shannon
rate-distortion function. From a practical perspective however the gap
to the rate-distortion bound decreases quickly in the degree.

\begin{example}[Generator-Regular LDGM Codes]
\label{exa:rdgeneratorregular}
Consider codes with generator degree equal to $\dl$ and
an arbitrary degree distribution on the check nodes.
In this case we have $f(x)=1+x^\dl$ and 
$a(x) = \frac{\dl x^\dl}{1+x^\dl}$.
Figure~\ref{fig:rdgeneratorregular} compares the lower bound to
the rate-distortion curve for $\dl=1$, $2$, and $3$. For each case the
achievable region is strictly bounded away from the Shannon rate-distortion curve.
\begin{figure}[htp]
\begin{center}
\input{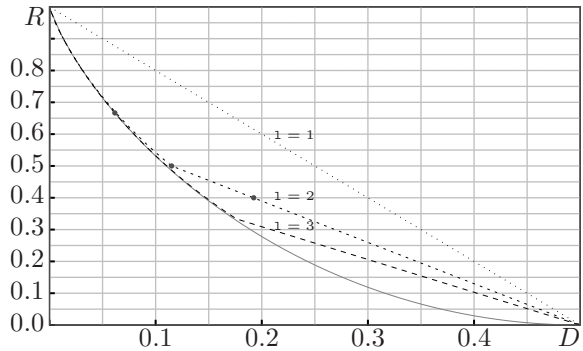}
\end{center}
\caption{\label{fig:rdgeneratorregular}
Bounds for $L(x)=x^\dl$ for $\dl=1$, $2$, and $3$.
For $\dl=2$ the $3$ gray dots correspond to the special
cases $R=\frac23$, $R=\frac12$, and $R=\frac25$ respectively.
The corresponding lower bounds on the distortion are
$D(\frac23) \geq 0.0616> 0.0614905$ (rate-distortion bound), 
$D(\frac12) \geq0.115 > 0.11$ (rate-distortion bound), and $D(\frac25) \geq 0.1924 >0.1461$ (rate-distortion bound).
}
\end{figure}
\end{example}

\begin{example}[$(\dl, \dr)$-Regular LDGM Codes]
In this case we have $R=\dl/\dr$ and $L(x)=x^{\dl}$.
The same bound as in Example~\ref{exa:rdgeneratorregular} applies.
The three special cases $(\dl=2, \dr=3)$, $(\dl=2, \dr=4)$, and $(\dl=2, \dr=5)$,
which correspond to $R=\frac23$, $R=\frac12$, and $R=\frac25$ respectively,
are marked in Figure~\ref{fig:rdgeneratorregular} as gray dots.
\end{example}

\begin{example}[$\dr$-Regular LDGM Codes of Rate $R$]
Assume that all check nodes have degree $\dr$ and that the 
connections are chosen uniformly at random with repetitions.
For large blocklengths this implies that the degree distribution
on the variable nodes converges to a Poisson distribution, i.e., we have
in the limit
\begin{align*}
L(x) & = \sum_{i=1}^{\infty} L_i x^i = e^{\frac{\dr}{R} (x-1)}.
\end{align*}
Let us evaluate our bound for this generator degree distribution.
Note that since the average degree of the {\em check} nodes is fixed we have a different
generator degree distribution $L(x)$ for each rate $R$.
Figure~\ref{fig:rdsourceregular} compares the resulting bound with the Shannon rate-distortion function
as well as the bound of Theorem~\ref{the:dwrbound}. The new bound
is slightly tighter. But more importantly, it applies to {\em any} LDGM code.  
\begin{figure}[htp]
\begin{center}
\input{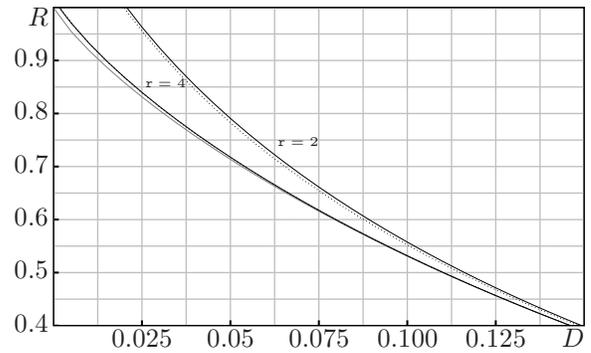}
\end{center}
\caption{\label{fig:rdsourceregular}
Lower bound on achievable $(R, D)$ pairs for $\dr$-regular LDGM codes with a
Poisson generator degree distribution and $\dr=2, 4$. The dashed curve corresponds to 
the bound of Theorem~\ref{the:dwrbound} and the solid black curve represents the bound
of Theorem~\ref{the:boundviacounting}.
The gray curve is the Shannon rate-distortion tradeoff. 
}
\end{figure}

\end{example}

{\em Proof of Theorem~\ref{the:boundviacounting}.}
From the statement in Theorem~\ref{the:boundviacounting} you see that the
bound consists of a portion of the curve $(\hat{D}(x), \hat{R}(x))$
and a straight-line portion.  The straight-line portion is easily
explained. Assume that all generator nodes have degree $\dl$ (for the
general case replace all mentions of $\dl$ by the average degree
$L'$). Then the maximum number of check nodes that can depend on the
choice of generator nodes is $n \dl$. Therefore, if the rate $R$ is
lower than $\frac{1}{\dl}$ then at least a fraction $(1-R \dl)$ of the
check nodes cannot be connected to any generator node. For those nodes
the average distortion is $\frac12$, whereas for the fraction $R \dl$
of the check nodes which are (potentially) connected to at least one
generator node the best achievable distortion is the same for any $0 \leq
R \leq \frac{1}{\dl}$.  It suffices therefore to restrict our attention
to rates in the range $[\frac{1}{L'}, 1]$ and to prove that their $(R,
D)$ pairs are lower bounded by the curve $(\hat{D}(x), \hat{R}(x))$.

As a second simplification note that although the bound is valid for
all blocklengths $m$ we only need to prove it for the limit of infinite
blocklengths.  To see this, consider a particular code of blocklength
$m$. Take $k$ identical copies of this code and consider these $k$
copies as one code of blocklength $k m$. Clearly, this large code has
the same rate $R$, the same generator degree distribution $L(x)$,
and the same distortion $D$ as each component code. By letting
$k$ tend to infinity we can construct an arbitrarily large code of
the same characteristics and apply the bound to this limit.  Since our bound
below is valid for {\em any} sequence of codes whose blocklength tends to
infinity the claim follows.

Pick $w \in \naturals$ so that $D m + w \leq \frac{m}{2}$. Then
\begin{align*}
|{\mathcal C}(D)| 
& = |\bigcup_{\hat{s} \in \code} {\mathcal B}(\hat{s}, D m)| \\
& \stackrel{\text{(i)}}{\leq}  \frac{1}{A_m(w)} \sum_{\hat{s} \in \code} |{\mathcal B}(\hat{s}, Dm +w)| \\
& \stackrel{\text{(ii)}}{\leq} 2^{-mR \log \frac{f(x_{\omega})}{x_{\omega}^{\omega}}+o_m(1)} 2^{m R} 2^{m h(D+w/m)} \\
& \stackrel{\text{(iii)}}{=} 2^{m (-R \log \frac{f(x_{\omega})}{x_{\omega}^{a(x_{\omega})}}+ 
R+ h(D+a(x_{\omega}) R) + o_m(1))}.
\end{align*}

To see (i) note that a ``big'' sphere ${\mathcal B}(\hat{s}, Dm +w)$, where $\hat{s} \in \code$, contains 
all ``small'' spheres of the form ${\mathcal B}(\hat{s}', D m)$, where $\hat{s}' \in \code$ so that
$\distortion(\hat{s}, \hat{s}') \leq w$.
Let $A_m(w)$ be the number of codewords of Hamming weight at most $w$. 
Then, by symmetry, each small sphere
${\mathcal B}(\hat{s}', Dm)$ is in exactly $A_m(w)$ big spheres ${\mathcal B}(\hat{s}, Dm +w)$.
It follows that every point in $\bigcup_{\hat{s} \in \code} {\mathcal B}(\hat{s}, D m)$ is counted at least
$A_m(w)$ times in the expression $\sum_{\hat{s} \in \code} |{\mathcal B}(\hat{s}, Dm +w)|$.

Consider now step (ii). We need a lower bound on $A_m(w)$. Assume at first that
all generator nodes have degree $\dl$.  Assume that exactly $g$ generator nodes
are set to $1$ and that all other nodes are set to $0$. There are $\binom{mR}{g}$
ways of doing this. Now note that for each such constellation the weight
of the resulting codeword is at most $w=g \dl$. It follows that in the generator regular case
we have
\begin{align} \label{equ:anofwone}
A_m(w) \geq  \sum_{g=0}^{w/\dl} \binom{mR}{g}. 
\end{align}
We can rewrite (\ref{equ:anofwone}) in the form
\begin{align} \label{equ:anofwtwo}
A_m(w) & \geq  \sum_{i=0}^{w} \text{coef}\{(1+x^\dl)^{mR}, x^i\}, 
\end{align}
where $\text{coef}\{ (1+x^\dl)^{mR}, x^i\}$ indicates the coefficient of the polynomial
$(1+x^\dl)^{mR}$ in front of the monomial $x^i$. 
The expression (\ref{equ:anofwtwo}) stays valid also for irregular generator degree distributions $L(x)$
if we replace $(1+x^\dl)^{mR}$ with $f(x)^{mR}$, where $f(x)=\prod_i(1+x^i)^{L_i}$ as defined in
the statement of the theorem. This of course requires that $n$ is chosen in such a way
that $n L_i \in \naturals$ for all $i$. 

Define $N_m(w) = \sum_{i=0}^{w} \text{coef}\{f(x)^{mR}, x^i\}$, so that
(\ref{equ:anofwtwo}) can be restated as $A_m(w) \geq N_m(w)$.  Step (ii) now
follows by using the asymptotic expansion of $N_m(w)$ stated as Theorem~1
\cite{MB04}, where we define $\omega=w/(mR)$ and where $x_{\omega}$ is the
unique positive solution to $a(x)=\omega$.

Finally, to see (iii) we replace $w$ by $mR a(x_{\omega})$ and thus
we get the claim.
Since this bound is valid for any $w \in \naturals$ so that $D m + w \leq \frac{m}{2}$ we get the bound
\begin{align*}
\lim_{m \rightarrow \infty} \frac{1}{m} \log |{\mathcal C}(D)| \leq g(D, R),
\end{align*}
where
\begin{align*}
g(D, R) & = \inf_{\stackrel{x \geq 0}{D+a(x)R \leq \frac12}} -R \log \frac{f(x)}{x^{a(x)}}+ R+ h(D+a(x) R).
\end{align*}

Now note that as long as $g(D, R)<1$,
$|{\mathcal C}(D)|$ is exponentially small compared to $2^m$.
Therefore, looking back at (\ref{equ:averagedistortion}) we see that in this case
the average distortion converges to at least $D$ in the limit $m \rightarrow \infty$.
We get the tightest bound by looking for the condition for equality, i.e. by looking
at the equation $g(R, D)=1$. 
If we take the derivative with respect to $x$ and set it to $0$ then we get the condition
\begin{align*}
\frac{x}{1+x} = D+R a(x).
\end{align*}
Recall that $D + a(x) R \leq \frac12$, so that this translates to $x \leq 1$.
This means that $x \leq 1$. Replace $D+a(x) R$ in the entropy term by $\frac{x}{1+x}$,
set the resulting expression for $g(R, x)$ equal to $1$, and solve for $R$.
This gives $R$ as a function of $x$ and so we also get $D$ as a function
of $x$. We have
\begin{align*}
R(x) = \frac{1-h(\frac{x}{1+x})}{1-\log \frac{f(x)}{x^{a(x)}}}, \,\,
D(x) = \frac{x}{1+x} - a(x) R(x) .
\end{align*}
A check shows that $x=0$ corresponds to $(D, R)=(0, 1)$ and that $x=1$ corresponds to 
$(D, R)=(\frac{L'-1}{2 L'}, \frac{1}{(L')^2})$. Further, $R$ and $D$ are monotone functions of $x$.
Recall that we are only interested in the bound for $R \in [\frac{1}{L'}, 1]$. We get the corresponding
curve by letting $x$ take values in $[0, x(\frac{1}{L'})]$. For smaller values of the rate
we get the aforementioned straight-line bound.

Looking at the above expression for $g(D, R)$ one can see why this
bound is strictly better than the rate-distortion curve for $D \in (0,
\frac12)$. Assume at first that the generator degree distribution is
regular. Let the degree be $\dl$. In this case a quick check shows
that $-R \log \frac{f(x)}{x^{a(x)}}$ is equal to $-R h(\frac{a(x)}{\dl})$. Since
$a(0)=0$ we get the rate distortion bound if we set $x=0$.
The claim follows by observing that $a(x)$ is a continuous strictly
increasing function and that $h(x)$ has an infinite derivative at $x=0$
while $h(D+a(x)R)$ has a finite derivative at $x=0$.  It follows that
there exists a sufficiently small $x$ so that $R h(\frac{a(x)}{\dl})$ is strictly
larger than $h(D+a(x)R)-h(D)$ and so that $D+a(x) R \leq \frac12$.  Hence, $g(D,
R)$ is strictly decreasing as a function of $x$ at $x=0$. This bounds
the achievable distortion strictly away from the rate-distortion bound.
The same argument applies to an irregular generator degree distribution;
the simplest way to see this is to replace $\dl$ by the maximum degree
of $L(x)$.

\section{Bound Via Test Channel}\label{sec:boundviatestchannel}
\begin{figure}[htp]
\begin{center}
\input{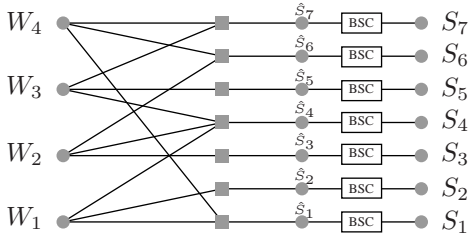}
\end{center}
\caption{\label{fig:testchannel} 
The generator words $W$ are chosen uniformly at random from ${\mathcal W}$.
This generates a codeword $\hat{S}$ uniformly at random. Each component of
$\hat{S}$ is then sent over a binary symmetric channel with transition probability $D'$.
}
\end{figure}

Instead of using a combinatorial approach to bound $|{\mathcal C}(D)|$
one can also use a probabilistic argument using  the ``test channel''
shown in Figure~\ref{fig:testchannel}.

For the cases we have checked 
the resulting bound is numerically identical to the bound of
Theorem~\ref{the:boundviacounting} (excluding the straight-line portion).
We restrict our exposition to the regular case. 
The generalization to the irregular case is straightforward.
\begin{theorem}[Bound Via Test Channel]\label{the:boundviatestchannel}
Let $\code$ be an LDGM code with blocklength $m$, 
generator degree distribution $L(x)=x^{\dl}$, and rate $R$.
Then for any pair $(R, D)$, where $D$ is the average distortion, we have
\begin{align*}
R & \geq \sup_{D \leq D' \leq \frac12} \frac{1-h(D)-\text{KL}(D \| D')}{1-\log_2\Bigl(1+\frac{(D')^\dl}{(1-D')^\dl} \Bigr)} \\
& \geq \frac{1-h(D)}{1-\log_2\Bigl(1+\frac{D^\dl}{(1-D)^\dl} \Bigr)} > 1-h(D),
\end{align*}
where $\text{KL}(D \| D')=D \log_2(D/D')+(1-D) \log_2((1-D)/(1-D'))$.
\end{theorem}
{\em Proof.} 
The same remark as in the proof of Theorem~\ref{the:boundviacounting} applies: although
the bound is valid for any blocklength it suffices to prove it for the limit of
blocklengths tending to infinity. Also, for simplicity we have not stated the bound
in its strengthened form which includes a straight-line portion. But the same technique
that was applied in the proof of Theorem~\ref{the:boundviacounting} applies also to the present case.

As remarked earlier, the idea of the proof is based on bounding
$|{\mathcal C}(D)|$ by using the ``test channel.'' More precisely,
choose $W$ uniformly at random from the set of
all binary sequences of length $m R$.  Subsequently compute
$\hat{S}$ via $\hat{S} = W G$, where $G$ is the generator matrix
of the LDGM code. Finally, let $S=\hat{S}+Z$, where $Z$ has iid components with
$\prob\{Z_i=1\}=D'$.

Consider the set of sequences $s \in {\mathcal C}(D)$.
For each such $s$ we know that there exists an $\hat{s} \in \code$ so
that $\distortion(s, \hat{s}) \leq Dm$.  
We have
\begin{align*}
& \prob\{S=s \mid s \in {\mathcal C}(D) \} \\
& =  \sum_{\hat{s}' \in \code} \prob\{S=s, \hat{S}=\hat{s}' \mid s \in {\mathcal C}(D) \} \\ 
& = \sum_{w=0}^{m} \sum_{\hat{s}' \in \code: \distortion(\hat{s}', \hat{s})=w} \prob\{S=s, \hat{S}=\hat{s}' \mid s \in {\mathcal C}(D) \} \\ 
& = \sum_{w=0}^{m} A_{m}(w) \prob\{S=s, \hat{S}=\hat{s}' \mid s \in {\mathcal C}(D), \distortion(\hat{s}', \hat{s})=w \} \\ 
& =  \sum_{w=0}^{m} A_{m}(w) 2^{-m R} \Bigl(\frac{D'}{1-D'}\Bigr)^{\distortion(s, \hat{s}')} (1-D')^m
\end{align*}
\begin{align*}
& \geq  \sum_{w=0}^{m} A_{m}(w) 2^{-m R} \Bigl(\frac{D'}{1-D'}\Bigr)^{\distortion(s, \hat{s})+\distortion(\hat{s}, \hat{s}')} (1-D')^m \\
& \stackrel{ \distortion(\hat{s}', \hat{s})=w}{=}  \sum_{w=0}^{m} A_{m}(w) 2^{-m R} \Bigl(\frac{D'}{1-D'}\Bigr)^{\distortion(s, \hat{s})+w} (1-D')^m \\
& \stackrel{\distortion(s, \hat{s}) \leq D m}{\geq}  \sum_{w=0}^{m} A_{m}(w) 2^{-m R} \Bigl(\frac{D'}{1-D'}\Bigr)^{D m+w} (1-D')^m \\
& = 2^{-m R -mh(D)-m \text{KL}(D \| D')} \sum_{w=0}^{m} A_m(w) \Bigl(\frac{D'}{1-D'}\Bigr)^{w},
\end{align*}
where $A_m(w)$ denotes the number of codewords in $\code$ of Hamming weight $w$. Due to
the linearity of the code this is also the number of codewords in $\code$ of Hamming
distance $w$ from $\hat{s}$.
Using summation by
parts and setting $c=D'/(1-D')<1$,  we have
\begin{align*}
& \sum_{w=0}^{m} A_m(w) c^w \\
& = c^{m+1} 2^{mR}+ \sum_{w=0}^{m}\Bigl(\sum_{i=0}^{w-1} A_m(i) \Bigr) (c^w-c^{w+1}) \\
& \stackrel{(\ref{equ:anofwtwo})}{\geq} c^{m+1} 2^{mR}+ \sum_{w=0}^{m}\Bigl(\sum_{i=0}^{\lfloor(w-1)/\dl \rfloor} 
\binom{mR}{i} \Bigr) (c^w-c^{w+1}) \\
& = \sum_{w=0}^{\lfloor m/\dl \rfloor} \binom{mR}{w} c^{\dl w} + c^{m+1} \Bigl(2^{mR} -
\sum_{i=0}^{\lfloor m/\dl \rfloor} \binom{mR}{i} \Bigr) \\
& \geq \sum_{w=0}^{\lfloor m/\dl \rfloor} \binom{mR}{w} c^{\dl w} \geq \frac1m (1+c^\dl)^{m R}.
\end{align*}
The last step is valid as long as $\frac{R c^\dl}{1+c^\dl} < \frac{1}{\dl}$. In
this case the maximum term (which appears at $\frac{R c^\dl}{1+c^\dl} m$) is
included in the sum (which goes to $m/\dl$) and is thus greater than equal to
the average of all the terms, which is $\frac1m (1+c^\dl)^{m R} $ . This
condition is trivially fulfilled for $R \dl < 1$. Assume for a moment that it
is also fulfilled for $R \dl \geq 1$ and the optimum choice of $D'$.  It then
follows that
\begin{align*}
\prob\{S=s \mid s \in {\mathcal C}(D) \} & 
\geq \frac1m 2^{-m(R+h(D)+\text{KL}(D \| D')-R \log_2(1+c^\dl))}.
\end{align*}
Since 
\begin{align*}
1 & = \sum_{s \in \GF^m} \prob\{S=s\} 
\geq \sum_{s \in {\mathcal C}(D)} \prob\{S=s\} \\
& \geq |{\mathcal C}(D)| \frac1m 2^{-m(R+h(D)+\text{KL}(D \| D')-R \log_2(1+c^\dl))},
\end{align*}
we have
$|{\mathcal C}(D)| \leq m 2^{m(R+h(D)+\text{KL}(D \| D')-R \log_2(1+c^\dl))}$.
Proceeding as in (\ref{equ:averagedistortion}), we have
\begin{align*}
& \expectation[\distortion(S, {\decoder}({\encoder}(S)))] 
\geq D \bigl(1-2^{-m} |{\mathcal C}(D)| \bigr) \\
& \geq D \bigl(1-m 2^{m(R+h(D)+\text{KL}(D \| D')-R \log_2(1+c^\dl)-1)}  \bigr).
\end{align*}
We conclude that if for some $D \leq D' \leq \frac12$, 
$R+h(D)+\text{KL}(D \| D')-R \log_2(1+\frac{(D')^\dl}{(1-D')^\dl})-1<0$
then the distortion is at least $D$. All this is still conditioned on
$\frac{R \dl c^\dl}{1+c^\dl} < 1$ for the optimum choice of $D'$.
For $R \dl <1$ we already checked this. So assume that $R \dl \geq 1$.
The above condition can then equivalently be written
as $D' < \frac{1}{1+(R \dl -1)^{\frac{1}{\dl}}}$.
On the other hand, taking the derivative of our final expression on
the rate-distortion function with respect to $D'$ we get the condition
for the maximum to be
$D' = \frac{1}{1+(1+\frac{R \dl}{D'-D})^{\frac{1}{\dl}}} < \frac{1}{1+(R \dl -1)^{\frac{1}{\dl}}}$.
We see therefore that our assumption $\frac{R \dl c^\dl}{1+c^\dl} < 1$ is also correct
in the case $R \dl \geq 1$.

Numerical experiments show that the present bound yields for the regular case
identical results as plotting the curve corresponding to $g(D, R)=1$, where
$g(D, R)$ was defined in the proof of Theorem~\ref{the:boundviacounting}.
This can be interpreted as follows. Choose $D'$ equal to the optimal radius of the Hamming
ball in the proof of Theorem~\ref{the:boundviacounting}. Then the points $\hat{s}'$ that
contribute most to the probability of $S=s$ must be those that have a distance to $\hat{s}$ of
$m(D'-D)$.

\section{Discussion and Open Questions}\label{sec:openquestions}
In the preceding sections we gave two bounds. Both of them are
based on the idea of counting the number of points that are ``covered''
by spheres centered around the codewords of an LDGM code. In the 
first case we derived a bound by double counting this number. In the second
case we derived a bound by looking at a probabilistic model using the test channel.

An interesting open question is to determine the exact relationship of
the test channel model to the rate-distortion problem.
More precisely, it is tempting to conjecture that a pair $(R, D)$ is only achievable
if $H(S)=m$ in this test channel model. This would require to show
that only elements of the typical set of ${\mathcal S}$ under the test channel model
are covered, i.e., have code words within distance $D$. For the test channel model it is 
very easy to determine a criterion in the spirit of Gallager's original bound.
We have
\begin{align*}
H(S) 
& = H(W)+H(S \mid W)- H(W \mid S) \\
& = m R + m h(D) - \sum_{g=1}^{m R} H(W_g \mid S, W_{1}, \dots, W_{g-1}) \\
& \stackrel{\text{(i)}}{\leq} m R + m h(D) - \sum_{g=1}^{m R} H(W_g \mid S, W_{\sim g}) \\
& \stackrel{\text{(ii)}}{=} m R + m h(D) - \sum_{g=1}^{m R} H(W_g \mid S_g, W_{\sim g}),
\end{align*}
where $S_g$ denotes the subset of the components of the $S$ vectors which
are connected to the generator $g$.
Step (i) follows since conditioning decreases entropy. Step (ii) follows since
knowing ($S_g, W_{\sim g}$), $W_g$ is not dependent on $S_{\sim g}$. The term
$H(W_g \mid S_g, W_{\sim g}) $  represents the EXIT function of a repetition
code when transmitting over BSC($D$) channel. 
If one could show that $H(S)=m$ is a necessary condition for 
achieving average distortion of $D$ then a quick calculation shows that
the resulting bound would read
\begin{align*}
R &  \geq 
\frac{1 - h(D)}{1-\sum_{i=0}^{\dl} \binom{\dl}{i} (1-D)^i 
D^{\dl-i} \log_2\Bigl(1+\bigl(\frac{D}{1-D}\bigr)^{2 i -\dl}\Bigr)}. 
\end{align*}
This ``bound'' is similar in spirit to the original bound given by Gallager, except
that in Gallager's original bound for LDPC codes we have a term corresponding to the
entropy of single-parity check codes, whereas here we have terms that correspond
to the entropy of repetition codes; this would be quite fitting given the duality of the problems.

\section*{Acknowledgment} We gratefully acknowledge the support by the
Swiss National Science Foundation under grant number 200020-113412.

\bibliographystyle{IEEEtran}
\bibliography{lth,lthpub}

\newcommand{\SortNoop}[1]{}
\begin{thebibliography}{10}
\providecommand{\url}[1]{#1}
\csname url@rmstyle\endcsname
\providecommand{\newblock}{\relax}
\providecommand{\bibinfo}[2]{#2}
\providecommand\BIBentrySTDinterwordspacing{\spaceskip=0pt\relax}
\providecommand\BIBentryALTinterwordstretchfactor{4}
\providecommand\BIBentryALTinterwordspacing{\spaceskip=\fontdimen2\font plus
\BIBentryALTinterwordstretchfactor\fontdimen3\font minus
  \fontdimen4\font\relax}
\providecommand\BIBforeignlanguage[2]{{%
\expandafter\ifx\csname l@#1\endcsname\relax
\typeout{** WARNING: IEEEtran.bst: No hyphenation pattern has been}%
\typeout{** loaded for the language `#1'. Using the pattern for}%
\typeout{** the default language instead.}%
\else
\language=\csname l@#1\endcsname
\fi
#2}}

\bibitem{MaYe03}
E.~Martinian and J.~Yedidia, ``Iterative quantization using codes on graphs,''
  in \emph{Proc. of the Allerton Conf. on Commun., Control, and Computing},
  Oct. 2003.

\bibitem{CiMe05}
S.~Ciliberti and M.~Mezard, ``The theoretical capacity of the parity source
  coder,'' \emph{J. Stat. Mech.}, 2005.

\bibitem{CiMeZe05}
S.~Ciliberti, M.~Mezard, and R.~Zecchina, ``Lossy data compression with random
  gates,'' \emph{Phys. Rev. Lett.}, vol.~95, 2005.

\bibitem{WaM05}
M.~J. Wainwright and E.~Maneva, ``Lossy source coding via message-passing and
  decimation over generalized codewords of {LDGM} codes,'' in \emph{Proc. of
  the IEEE Int. Symposium on Inform. Theory}, Adelaide, Australia, Sept. 2005,
  pp. 1493--1497.

\bibitem{BZ03}
A.~Braunstein and R.~Zecchina, ``Survey propagation as local equilibrium
  equations,'' \emph{J. Statistical Mechanics: Theory and Experiment}, June
  2004.

\bibitem{FiFr07}
T.~Filler and J.~Fridrich, ``Binary quantization using belief propagation with
  decimation over factor graphs of ldgm codes,'' in \emph{Proc. of the Allerton
  Conf. on Commun., Control, and Computing}, Sept. 2007.

\bibitem{MaWa06}
E.~Martinian and M.~J. Wainwright, ``Low-density codes achieve the
  rate-distortion bound,'' in \emph{Proc. of the Data Compression Conference},
  Snowbird, UT, Mar. 2006.

\bibitem{MaW06a}
------, ``Low-density constructions can achieve the {W}yner-{Z}iv and
  {G}elfand-{P}insker bounds,'' in \emph{Proc. of the IEEE Int. Symposium on
  Inform. Theory}, Seattle, WA, USA, July 2006, pp. 484--488.

\bibitem{MaW06b}
------, ``Analysis of {LDGM} and compound codes for lossy compression and
  binning,'' in \emph{Proc. of the IEEE Inform. Theory Workshop}, San Diego,
  CA, USA, Feb. 2006.

\bibitem{DiWaRa07}
A.~Dimakis, M.~Wainwright, and K.~Ramchandran, ``Lower bounds on the
  rate-distortion function of ldgm codes,'' in \emph{Proc. of the IEEE Inform.
  Theory Workshop}, 2007.

\bibitem{CoT91}
T.~M. Cover and J.~A. Thomas, \emph{Elements of Information Theory}.\hskip 1em
  plus 0.5em minus 0.4em\relax New York, NY, USA: Wiley, 1991.

\bibitem{MB04}
D.~Burshtein and G.~Miller, ``Asymptotic enumeration methods for analyzing
  {LDPC} codes,'' \emph{IEEE Trans. Inform. Theory}, vol.~50, no.~6, pp.
  1115--1131, June 2004.

\end{thebibliography}

\end{document}